\documentclass[lettersize,journal]{IEEEtran}
\usepackage{amsmath,amsfonts}
\usepackage{algorithmic}
\usepackage{algorithm}
\usepackage{array}
\usepackage[caption=false,font=normalsize,labelfont=sf,textfont=sf]{subfig}
\usepackage{textcomp}
\usepackage{stfloats}
\usepackage{url}
\usepackage{verbatim}
\usepackage{graphicx}
\usepackage{cite}
\usepackage{booktabs}
\usepackage{multirow}
\begin{document}

\title{Robust Calibrate Proxy Loss for Deep Metric Learning}

\author{
  Xinyue~Li, Jian~Wang, Wei~Song*, Yanling~Du, Zhixiang~Liu

  % <-this % stops a space
  \thanks{This work was supported in part by the National Natural Science Foundation of China under Grant 61972240, and Science and Technology Commission of Shanghai Municipality under Grant 20050501900.}
  % <-this % stops 
  \thanks{Xinyue~Li, Jian~Wang, Wei~Song, Yanling~Du and Zhixiang~Liu are with the Institute of Digital Ocean, Shanghai Ocean University, Shanghai 201306, China. (e-mail: wsong@shou.edu.cn).}% <-this % stops a space
}

\markboth{Journal of \LaTeX\ Class Files,~Vol.~14, No.~8, August~2021}%
{Shell \MakeLowercase{\textit{et al.}}: A Sample Article Using IEEEtran.cls for IEEE Journals}


\maketitle

\begin{abstract}
  The mainstream researche in deep metric learning can be divided into two genres: proxy-based and pair-based methods. Proxy-based methods have attracted extensive attention due to the lower training complexity and fast network convergence.
  However, these methods have limitations as the poxy optimization is done by network, which makes it challenging for the proxy to accurately represent the feature distrubtion of the real class of data.
  In this paper, we propose a Calibrate Proxy (CP) structure, which uses the real sample information to improve the similarity calculation in proxy-based loss and introduces a calibration loss to constraint the proxy optimization towards the center of the class features.
  At the same time, we set a small number of proxies for each class to alleviate the impact of intra-class differences on retrieval performance.
  % In particular, our method shows strong robustness against label noises.
  The effectiveness of our method is evaluated by extensive experiments on three public datasets and multiple synthetic label-noise datasets.
  The results show that our approach can effectively improve the performance of commonly used proxy-based losses on both regular and noisy datasets.
  % The results show that our Calibrate Proxy loss (CP loss) achieves good performance on both clean and noisy datasets.
  % The results show that our Calibrate Proxy loss (CP loss) achieves good performance on both clean and noisy datasets, and can compete with SOTA methods.
\end{abstract}


\section{Introduction}

\IEEEPARstart{D}{eep} metric learning is fundamental research, which is widely used in various fields, such as image retrieval \cite{1-1-retrieval,1-2-retrieval} and classification \cite{2-1-image_classification,2-2-image_classification}, cross-modal matching \cite{3-1-cross-modal_matching,3-2-cross-modal_matching}, person re-identification \cite{4-1-reid,4-2-reid_10}, etc.
It can also provide solutions for many practical areas \cite{5-1-practical_areas,5-2-practical_areas,5-3-practical_areas}.
Metric learning is originated from the distance metric, and gradually evolved into deep metric learning with the development of deep neural networks.
The objective of deep metric learning is to train the network to learn a reasonable embedding space, where similar samples are close to each other and dissimilar samples are separated, to effectively measure the similarity between samples.

The mainstream deep metric learning can be divided into two categories: pair-based \cite{7_23,8,11,12,tmmRS} and proxy-based \cite{15,17_22_26,19,18} methods.
The pair-based methods calculate the similarity of pairs, which can extract the sample features with rich information.
Due to the high training complexity, they often cooperate with sampling techniques \cite{6-1-sampling,6-2-sampling,6-3-sampling} and are not robust to noise data.
The proxy-based methods use a proxy to represent the features of a class and calculate the similarity between samples and proxies.

\begin{figure}
  \centering
  \includegraphics [width=0.4\textwidth]{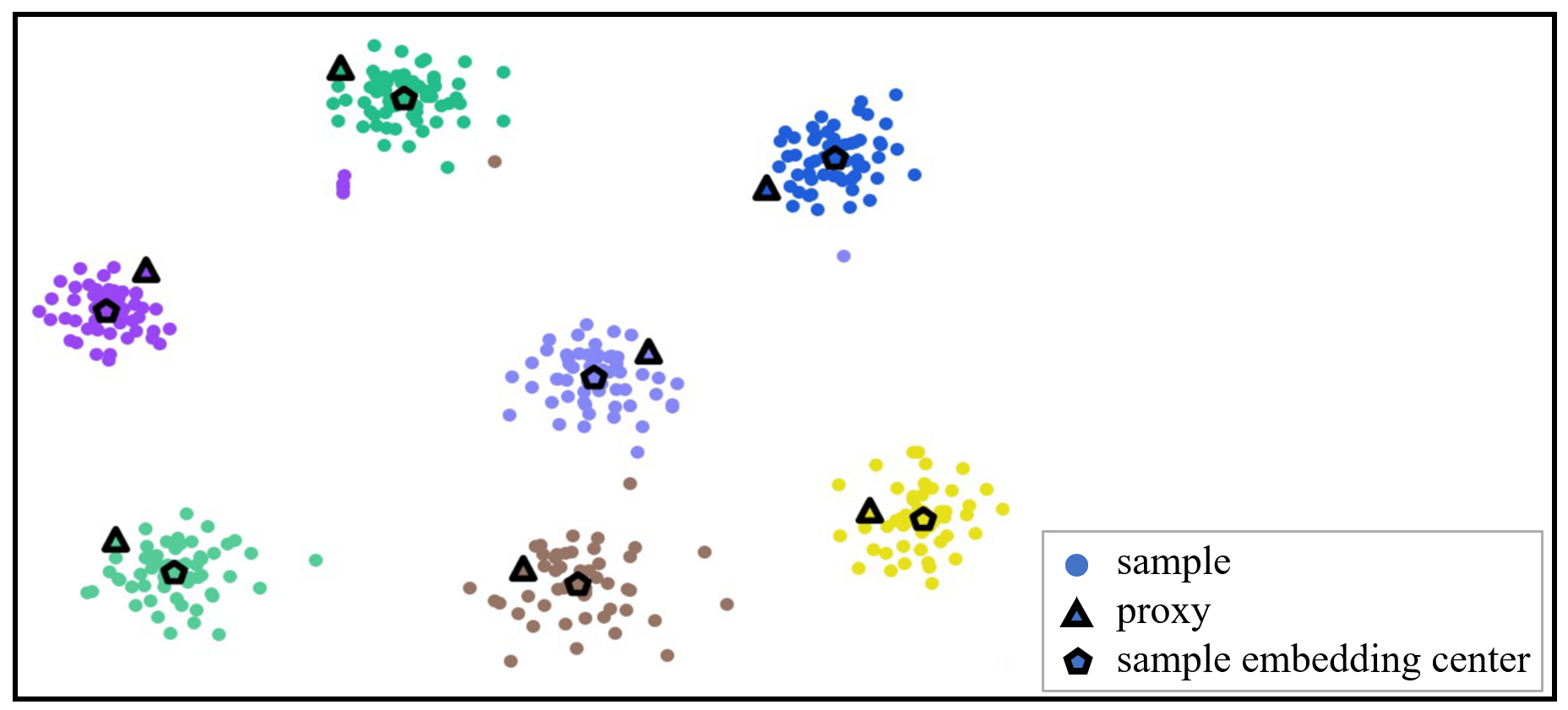}
  \caption{The t-SNE visualization of the embedding distribution of data samples, proxies (triangles) and sample embedding centers (pentagons) for the model trained with the Proxy Anchor loss on CUB-200-2011 dataset. The figure shows there is obvious deviation between the proxy and the sample embedding center for each class.}
  \label{fig-tsne}
\end{figure}

Compared with the pair-based methods, the proxy-based methods can effectively reduce the training complexity and accelerate network convergence, and have stronger robustness to noise data.
Since the proxies are randomly initialized, the embedding vectors of the samples and proxies deviate greatly. In the training process, they both optimized in their respective embedding spaces, so there is an unbridgeable deviation between proxies and the real sample distribution.
We present a t-SNE visualization of the traditional proxy-based method in Fig.\ref{fig-tsne} to illustrate this phenomenon.
Our intuition is that pair-based methods and proxy-based methods are partially complementary.
% However, the proxy optimization relies on the network, and there is an embedding deviation with real sample distribution.
% We present a t-SNE visualization of the traditional proxy-based method in Fig.\ref{fig-tsne} to illustrate this phenomenon.
% Our intuition is that pair-based methods and proxy-based methods are partially complementary.
Therefore, on the basis of the proxy-based method, we use the real data information to calibrate the proxy towards the actual class center.

We studied many datasets and found that samples within the same class have obvious feature differences, and these differences show regular aggregation,
e.g., birds can be divided into the flying state and the standing state, and the same object can differ greatly due to different shooting angles.
This phenomenon also exists in the real world. Therefore, we consider that setting multiple proxies for each class may more comprehensively express the diverse feature differences in a class.

% In addition, given that the acquisition and labeling of datasets is always a complicated task, most of the data in real-world applications are not pure enough and mixed with destructive labeling noise.
% Therefore, it is of more practical significance to enhance the noise robustness of the algorithm and improve the performance on noisy datasets.

In summary, the objective of our study is to design a proxy-based method that enables the proxy to represent real class features.
First, we propose the Calibrate Proxy structure. In this structure, a global embedding center (short for global center) is designed to express class features together with the typical proxy. We use the similarity between the global center and samples to improve the loss calculation, and use the deviation between the global center and the proxies to imposes a constraint to calibrate the proxy optimization towards the global center.
Second, we assign multiple proxies to each class to represent the variation of intra-class features, which has a positive impact on improving retrieval performance.
In particular, our Calibrate Proxy structure and Multi-proxy module are not limited to combining with a specific proxy-based loss.
% The above improvements are not limited to a particular proxy-based loss, so we apply the proposed CP structure to some commonly used proxy-based losses and jointly improve their performance.
% In particular, our method performs better than other methods in the noise-resistant deep metric learning field without performing specific noise handling procedures.
% In particular, our method also shows good noise resistance.

% In summary, the objective of our study is to design a proxy-based method that enables the proxy to represent real class features and improve its robustness against label noise.
% First, we propose the Calibrate Proxy structure. Specifically, we design a global embedding center (short for global center) to express class features together with the typical proxy.
% The global center is used to calculate the similarity with samples to improve the loss calculation.
% Then we add a constraint to calibrate proxy optimization toward the global center.
% Second, we assign multiple proxies to each class to represent the variation of intra-class features, which has a positive impact on improving retrieval performance.
% In particular, we validate our algorithm on image retrieval datasets with different noise ratios.
% Our method performs better than other methods in the noise-resistant deep metric learning field without performing specific noise handling procedures.

The main contributions of this work are summarized as follows:

1. We propose a Calibrate Proxy structure that combines the proxy center with the global center in order to calibrate the proxy optimization towards the real class center.

2. We assign multiple proxies to each class to deal with the variation and regular aggregation of features within the same class, and to optimize the samples toward a closer proxy in the same class.

% 3. We conduct extensive experiments on three widely-used datasets to validate the performance of our method, as well as experiments on synthetic label-noisy datasets with different noise ratios to validate the noise-resistance performance.
% The results show that our method achieves new performance on both regular and noisy datasets.
% The results show that our method reaches 69.8\% at Recall@1 and 28.34\% at MAP@R on the CUB-200-2011 dataset.
3. We combine the Calibrate Proxy structure and multi-proxy module in commonly used proxy-based losses and conduct extensive experiments on three widely-used datasets. The results show that our method achieves good performance on both regular and noisy datasets.

\section{Related works}

\subsection{Pair-based deep metric learning}

The early pair-based method is Contrastive loss \cite{7_23}. It consists of samples of the same class into positive pairs, and different classes of samples into negative pairs.
It requires positive pairs to be close to each other and negative pairs to be far away from each other.

Triplet loss \cite{8} takes each sample as a query sample, and selects samples from the same class and different classes to form a triplet.
It requires the distance between negative pairs in the triplet to be greater than that between positive pairs.
Many subsequent studies are carried out on the basis of triplet, such as N-pair loss \cite{9_20}, Beyond Triplet loss \cite{4-2-reid_10}, etc.

Lifted structure loss \cite{11} make full use of all the samples in mini-batch and mine the features of all pairs. It does not distinguish between query samples or positive samples. forcing the distance between each sample in the pair and all other negative samples to be greater than the given threshold.

Multi-similarity loss \cite{12} can fully utilize both positive and negative pairs by a more generalized weighting strategy.
Concretely, Multi-similarity loss mines informative pairs by the triplet criterion for both positive and negative pairs, then respectively assigns different weights for them.

Circle loss \cite{13} use a unified form, which is compatible with both class-level labels and pair-wise labels.
The author believes that pair-based loss is inflexible for a pair to have the same penalty strength.
Therefore, they weighted separately on the minimum unit pair to obtain a flexible optimization and a clear convergence objective.

Considering that the previous pair-based loss is limited by the size of the batch, Wang et al. \cite{14} proposeed Cross Batch Memory module and defined a slow drift phenomenon.
They prove that reaching a certain number of training times, the change of instance embeddings tends to be slow.
Thus, storing the history embedding of samples can be used as a complement to the current batch.

The pair-based method combines samples into pairs to calculate the loss, which can obtain rich sample information.
However, the disadvantages are higher training complexity, slower network convergence, and more sensitivity to label noise.

\subsection{Proxy-based deep metric learning}

Proxy-NCA loss \cite{15} randomly initializes the same number of proxies of the number of classes and optimizes each proxy to represent a certain class through network training.
Proxy-NCA loss can promote samples to proxy of the same classes and away from proxies of different classes.

Following the Proxy-NCA loss, Proxy Anchor loss \cite{17_22_26} takes the proxy as the anchor point, and flexibly adjusts the optimization strength according to the sample-proxy similarity.

Manifold Proxy loss \cite{19} extends N-pair loss to a form of proxy-based method.

Some methods assign more than one proxy or center to each class in order to enhance the representation ability of the proxy.
For example, SoftTriple loss \cite{16} extends softmax loss to multiple centers of every class and merges adjacent centers to represent the class features.
Hierarchical Proxy-based Loss \cite{18} organizes the proxies into a hierarchical structure so as to learn shared information across classes.

To improve the network's ability to deal with unknown data, Proxy Synthesis \cite{27} simulate invisible classes in the training set by synthesizing new classes through semantic interpolations.

Smooth Proxy Anchor loss \cite{21} increases a confidence module to Proxy Anchor loss \cite{17_22_26}. The confidence module uses the multi-classification results as the confidence to balance the sample weighting and reduce the influence of noisy samples.

The proxy-based methods can reduce the training complexity and accelerate network convergence, but its proxy optimization mainly depends on the performance of the network, which limits the ability to express the features of real-data classes.

\section{Proposed Method}

In this section, we first review the Proxy Anchor loss \cite{17_22_26}.
Then we elaborate on the proposed Calibrate Proxy structure and the Multi-proxy module.
% Finally, we make a complete definition of CP loss.
% In particular, our Calibrate Proxy structure and Multi-proxy module are not limited to combining with a specific proxy-based loss.
Finally, we provide a guide for applying such improvements plug and play to existing proxy-based losses.

\subsection{Review of Proxy Anchor loss}
Proxy Anchor loss \cite{17_22_26} is a representative proxy-based loss, which assigns a proxy $p \in P$ to each class, and uses each proxy as an anchor to associate with each sample $x \in X$ in the batch.
$P$ is the set of all proxies, $P^+$ is the set of proxies of the class that exists in the current batch, which is called the positive proxies, and $P^-$ is the set of negative proxies, $P^- = P - P^+$.
For each proxy $p$, all sample $x$ in this batch can be divided into a positive set $X_{p}^+$, and a negative set $X_{p}^-$, $X_{p}^- = X - X_{p}^+$.
Proxy Anchor loss is defined as:
\begin{equation}
  \label{eq1}
  \begin{aligned}
    L_{pa}(X) & = \frac{1}{{| {{P^ + }} |}}\sum_{p \in {P^ + }} {\log ( {1 + \sum\limits_{x \in X_p^ + } {e^{ -\alpha ( {s(x, p) - \delta } )}} } )} \\
              & + \frac{1}{{| P^ - |}}\sum\limits_{p \in P^ -} {\log ( {1 + \sum\limits_{x \in X_p^ - } {e^{\alpha ( {s(x, p) + \delta } )}} } )}    \\
  \end{aligned}
\end{equation}
where $s(,)$ is the cosine similarity between two embedding vectors.
The $\delta$  and $\alpha$ are hyperparameters that denote margin and scaling factors.
For the proxy $p \in P^+$, the loss promotes the similar sample $x \in X_{p^+}$ to be close to $p$, and separates the dissimilar samples $x \in X_{p^-}$ from $p$, and the strength is determined by the similarity of $x$ and $p$.

The proxy-based method assigns one proxy to each class to express the class features, forcing the sample of the whole class to be compact and optimized around the proxy.
This kind of method has less time complexity and is more robust to noisy labels.
However, proxy optimization is performed by the network, which lead a deviation between the proxies and the real samples of the same class in the embedding space.

\subsection{Calibrate Proxy Structure}

\begin{figure}
  \includegraphics [width=0.47\textwidth]{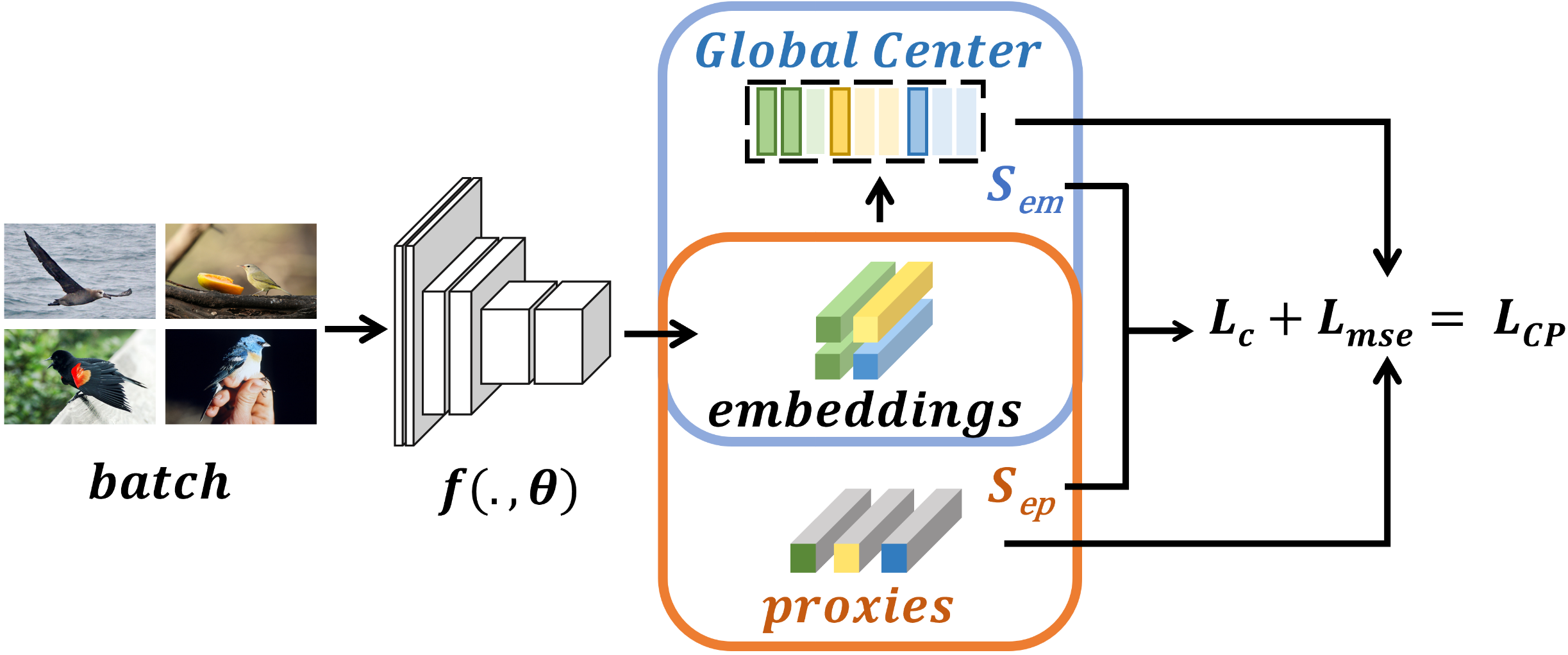}
  \caption{Calibrate Proxy Structure.
    % The samples in a batch enter the network $f(.,\theta)$ to get a set of embeddings.\\
    % The similarity between samples and proxies is denoted as $S_{ep}$ and the similarity between samples and global center is denoted as $S_{em}$, the loss $L_c$ is calculated by combining these two similarities.
    % The loss $L_{mse}$ is calculated by the proxies and global center, and calibrate proxy optimization towards the global center.
    % The final Calibrate Proxy loss is composed of $L_c$ and $L_{mse}$.
  }
  \label{fig_cp}
\end{figure}

We designed the Calibrate Proxy (CP) structure to eliminate the deviation between the proxy and sample embedding for the same class.
In the CP structure, a composite class center is proposed to accurately express class features, which consists of the proxy and a global center derived from the data samples.
We expect the global center can express the universal features of a real class.

As shown in Fig.\ref{fig_cp}, our proposed CP structure can be combined with a feature extraction network.
% The extracted feature vectors represent the embeddings expression of the samples and are processed with three components.

First, like the traditional proxy-based loss, we obtain a batch of sample embeddings through the network $f(.,\theta)$, and then calculate the similarity $S_{ep}$ between the sample embeddings and the proxies.

Second, we calculate the similarity $S_{em}$ between the samples and the global center, then combine the $S_{em}$ with $S_{ep}$ to calculate the loss $L_c$.
Moreover, we calculate the deviation between the proxies and the global center of the same class to obtain the constraint $L_{mse}$, so as to calibrate the optimization of the proxy towards the real class center.

Finally, we combine $L_c$ and $L_{mse}$ to get the calibrate proxy-based loss.
The following subsections explain the CP structure in three parts, global center, improved similarity calculation, and proxy calibration.

\subsubsection{Global Center}

We expect to find a set of embeddings derived from real samples as the global center for expressing class features. Here is a conflict.
To learn global embeddings of classes, we need more samples in each iteration.
However, to reduce the computational resource usage in network training, the capacity of the batch size is usually limited.
This makes the network perform local optimization based on the samples in the batch due to insufficient observation of the global embedding information.

To solve the problem, we implement the global center with a set of class queues, which store the embeddings of the samples previously entered.
On the one hand, it reduces the demand for computing resources caused by a large batch size.
On the other hand, storing more embeddings can be used to describe the overall data distribution for each class.
It is reasonable to use this global information to compensate for the deviation between the distribution of the proxy and the real class samples.

Let $Q$ denote the global center.
The sample embeddings are stored in $Q$ by class, $Q = \{Q_1, Q_2, \dots, Q_{n_c}\}$, where $n_c$ is the number of classes in the training set.
$n_q$ is the maximum number of samples in each class stored in the global center, and $Q_i$ denotes the queue of the $i$-th class, that is, $Q_i=\{q_{i1}, q_{i2}, \dots, q_{in_q} \}$.

When the number of samples in $Q_i$ exceeds the capacity $n_q$, the samples in $Q_i$ are updated in a first-in, first-out order to ensure the timeliness of the information in the global center.

\subsubsection{Improved Similarity Calculation}

We propose an improved similarity calculation that considers both the sample-global center similarity and the sample-proxy similarity, in order to obtain a more accurate indicator for loss calculation.

Specifically, $S_{ep}(x_i,P_c)$ denotes the similarity between sample $x_i$ and proxy $P_c$ for class $c$.
$S_{em}(x_i,B_c)$ denotes the similarity between sample $x_i$ and embeddings of class $c$ in the global center, corresponding to $S_{ep}$ and $S_{em}$ in Fig.\ref{fig_cp}, which are summed of $S_{cp}(x_i,C_c)$ for loss calculation.
The formula is as follows:
\begin{equation}
  \label{eq2}
  S_{cp}(x_i, C_c) = S_{em}(x_i,B_c) + S_{ep}(x_i, P_c)
\end{equation}
where $C$ denotes the composite class center jointly determined by the proxy and global center, $c$ denotes a certain class, and $S_{cp}(x_i,C_c)$ denotes the similarity between the $i$-th sample and the $c$-th class center.
The detail of $S_{em}(x_i,B_c)$ will describe in the following.

In the global center initialization, we learn from the experience of \cite{14} that there is a large feature drift in the early training period, so we set a start epoch $n_s$ for the global center.
When the training epoch is greater than $n_s$, we use all the sample information stored in the global center to calculate the similarity $S_{em}(x_i,B_c)$.
The formula is as follows:
\begin{equation}
  \label{eq3}
  S_{em}(x_i, B_c) = \frac{1}{n_q} \sum_{j=1}^{n_q}{S(x_i,b_{cj})}
\end{equation}
where $b_{cj}$ is the $j$-th sample of class $c$ in the global center, and $S_{em}(x_i,b_{cj})$ is the similarity between sample $x_i$ and $b_{cj}$.

The similarity $S_{cp}$ formed by $S_{em}$ and $S_{ep}$ can reasonably measure the similarity between samples and classes from two perspectives, the proxy center and the global center, to guide the correct optimization of the network.

\subsubsection{Proxy Calibration}

In this section, we use the global center to calibrate the proxy optimization.
The motivation for this method is that during the optimization, there is always a deviation between the samples and the proxies, and both of them are optimized in their respective embedding spaces.

This naturally reminds us of a problem in the cross-modal field: how to effectively eliminate feature misalignment in different modalities?
Numerous solutions point to adding constraints to the loss function to reduce the differences between samples of the same class between different modals, which can effectively narrow the heterogeneous gap \cite{26-1,26-2,26-3}.
The embedding gap between the same class proxies and real samples is similar to the heterogeneous gap cross modals.
Therefore, we introduce a constraint term $L_{mse}$ in the loss function and use the Mean Square Error (MSE) loss to calibrate the optimization of proxies toward the center of the sample embeddings, and we define as follows:
\begin{equation}
  \label{eq4}
  L_{mse}(P, B) = \sum_{c=1}^{n_c}{ \sum_{i=1}^{n_q}{(P_c - b_{ci})^2}}
\end{equation}
where $P_c$ is the proxy of class $c$, and $b_{ci}$ denotes the $i$-th sample embedding of the $c$-th class in the global center.

Specifically, we calculate the MSE loss of the proxy $P_c$ with all embeddings of the corresponding class in the global center to obtain the constraint $L_{mse}(P,B)$.
This loss calibrates the proxy embeddings to optimize toward the global center.

\subsection{Multi-proxy Module}

In this section, we introduce the Multi-proxy module to further improve the retrieval network performance by a certain number of proxies.

The practical basis for this idea is that samples of the same class in a dataset often show obvious feature differences, and such differences often cluster regularly.
We hope that assigning multiple proxies can effectively represent such data distribution, making each proxy better fit a cluster of sample features within the class.
We define the calculation of Multi-proxy as follows:
\begin{equation}
  \label{eq5}
  \begin{aligned}
    S_{ep}(x_i, P_c) & = S_{mul}(x_i, P_c)                                             \\
                     & = \sum_{j=1}^{n_p}{S(x_i, p_{cj})\cdot softmax(S(x_i, p_{cj}))} \\
  \end{aligned}
\end{equation}
where $n_p$ is the number of proxies assigned for each class, $p_{cj}$ is the $j$-th proxy of class $c$, $S(x_i,p_{cj})$ denotes the similarity between the $i$-th sample and $p_{cj}$.
We use the softmax function on $S(x_i,p_{cj})$, multiply it with itself and sum up to get the similarity $S_{ep}(x_i,P_c )$  between $x_i$ and all the proxies of class $c$, that is, the sample-proxy similarity $S_{ep}(x_i,P_c )$ in Eq.(\ref{eq2}).

Intuitively, we optimize the samples to be closer to proxies of the same class, which has a positive effect on the retrieval performance.
We emphasize that the Multi-proxy module does not significantly increase the time complexity of our method.
In the experiments part below, we validate the setting of the proxy number, and the results show that only a few proxies for each class can achieve good performance.

% \subsection{Calibrate Proxy Loss}
\subsection{Proxy-based Losses with Calibrate Proxy structure}

The proposed CP structure can be integrated with the existing proxy-based losses.
In the following, we provide a guide for applying the CP structure plug and play to existing proxy-based losses.
The CP structure consists of $L_c$ and $L_{mse}$, where $L_{mse}$ is the constraint of the proxy-global center, and the formula is as follows:
\begin{equation}
  \label{eq7}
  \begin{aligned}
    L_{cp} = L_{c} + \lambda L_{mse}
  \end{aligned}
\end{equation}
where $\lambda$ is used to adjust the influence of the loss $L_{mse}$.

Through the above analysis and the formulas (\ref{eq2}), (\ref{eq3}) and (\ref{eq5}), CP structure can be used as optimization item combined with multiple proxy-based losses as $L_c$ item.

% Through the above analysis and the formulas (\ref{eq2}), (\ref{eq3}) and (\ref{eq5}), our MHP method is shown in the form of optimization items as (\ref{eq:3}).

% In this section, we summarize the Calibrate Proxy structure and Multi-proxy module and finally design the Calibrate Proxy loss.
% Our CP loss consists of $L_c$ and $L_{mse}$, the $L_c$ loss is defined as follows:

\textbf{CP + Proxy Anchor Loss:}

We reviewed proxy anchor loss in Section 3.1, we rewrite it according to Calibrate Proxy Structure as Eq.(\ref{eq6}),

\begin{equation}
  \label{eq6}
  \begin{aligned}
    L_{PA}^{cp}(X) & = \frac{1}{{| {{C^ + }} |}}\sum_{c \in {C^ + }} {\log ( {1 + \sum\limits_{x \in X_c^ + } {e^{ -\alpha ( {S_{cp}(x, c) - \delta } )}} } )} \\
                   & + \frac{1}{{| C |}}\sum\limits_{c \in C} {\log ( {1 + \sum\limits_{x \in X_c^ - } {e^{\alpha ( {S_{cp}(x, c) + \delta } )}} } )}          \\
  \end{aligned}
\end{equation}
where $C$ is the set of all classes, $C^+$ is the set of classes present in the batch.
For each class $c \in C$, the sample set $X$ in batch can be partitioned into a positive sample set $X_{c}^+$ and a negative sample set $X_{c}^- = X_c - X_{c}^+$.
This part is consistent with the sample division method of Proxy Anchor loss, but we replace the proxy $p$ with a composite center $C_c$.
$S_{cp}(x,c)$ combines sample-proxy similarity and sample-global center similarity.
The $\alpha$ , $\delta$  in Eq.(\ref{eq6}) are hyperparameters.

\textbf{CP + Proxy NCA Loss:}

Proxy-NCA loss \cite{15} selects the anchor point $x$, the positive proxy with the same label as anchor, and the negative proxy set composed of all the remaining proxies to participate in the calculation of the loss.
We can rewrite it according to CP structure as (\ref{eq9}),

% \begin{equation}
%   \label{eq9}
%   L_{NCA}^{cp}({x,y,Z}) = -\log({\frac{{e^{{S_{x,y}^{cp}(x,y)}}}} {{\sum_{z \in Z} {e^{{S_{x,z}^{cp}(x,z)}}}} }} )
% \end{equation}

\begin{equation}
  \label{eq9}
  L_{NCA}^{cp}({x,c^+,C^-}) = -\log({\frac{{e^{{S_{x,c^+}^{cp}(x,c^+)}}}} {{\sum_{c \in C^-} {e^{{S_{x,c}^{cp}(x,c)}}}} }} )
\end{equation}
where $c^+$ is the positive class, and $C^-$ is the negative class set consisting of all remaining classes that participate in the calculation of loss.

\textbf{CP + SoftTriple Loss:}

SoftTriple loss\cite{16} associates samples $x$ with the centers of all classes, uses SoftMax to calculate the similarity between samples and a certain class, and finally computes the cross-entropy loss.
We change the optimization of SoftTriple loss from associating samples with proxies to associating with global centers, as shown in (\ref{eq10}),

% \begin{equation}
%   \label{eq10}
%   L_{soft}^{cp}(x) = -\log \frac{{e^{\lambda ( {S_{x,y}^{cp}(x,y) - \delta } )}}}{{e^{\lambda ( {S_{x,y}^{cp}(x,y) - \delta } )} + \sum\limits_{j \ne {y_i}} {e^{\lambda S_{x,j}^{cp}(x,j)}} }} \\
% \end{equation}

\begin{equation}
  \label{eq10}
  L_{soft}^{cp}(x) = -\log \frac{{e^{\lambda ( {S_{x,c^+}^{cp}(x,c^+) - \delta } )}}}{{e^{\lambda ( {S_{x,c^+}^{cp}(x,c^+) - \delta } )} + \sum\limits_{c \in C^-} {e^{\lambda S_{x,c}^{cp}(x,c)}} }} \\
\end{equation}

\section{Experiments}

% In this part, we first introduced the implementation details and datasets used in the experiments.
% Then, we conducted ablation experiments on the CUB-200-2011 \cite{28} and Cars-196 \cite{30} datasets.
% Finally, the proposed method was compared with the existing methods on three published datasets, which proved the effectiveness of our proposed method.
% In this part, we introduced the implementation details and datasets used in the experiments, then we conducted ablation experiments and compare our method with existing methods on three public datasets. 

% To fully evaluate the proposed Calibrate Proxy Loss, we follow the standardized evaluation protocol "MAP@R" recently proposed in "Reality Check" \cite{reality_check} and the traditional evaluation protocol "recall@k".

% In addition, we also conducted experiments on label-noise datasets to verify the robustness of our method against label-noise.

In this section, we first introduce the implementation details and the datasets used in the experiments. Then, we present the results of ablation experiments and the comparison results of our method with existing methods on three public datasets. Moreover, we verify the robustness of our method against label-noise in datasets.

For all the evaluations, we follow the standardized evaluation protocol ”MAP@R” recently proposed in "Reality Check" \cite{reality_check} and the traditional evaluation protocol "Recall@k".

\subsection{Implementation details}
We used the PyTorch framework to implement the CP structure.
Our experiments ran on a single Tesla P100 GPU with 16 GB RAM.
For a fair comparison experiment, we used batch normalized (BN)-Inception \cite{29} as the backbone feature extraction network and pre-trained on the ImageNet dataset\cite{RN64}.
And we set the embedding size to 512.
For all datasets, the input images were first resized to 256x256 and then cropped to 224x224.
Each model was trained for 60 epochs and the batch size was set to 150.
Input batches were randomly sampled during training.

\subsection{Datasets}

The experiment was run on three standard datasets, CUB-200-2011 \cite{28}, Cars-196 \cite{30}, Stanford Online Products (SOP) \cite{11}, and two noise datasets synthesized by CUB-200-2011 and Cars-196 with three different noise ratios.

CUB-200-2011 has 200 species of birds with 11,788 images.
We split 100 species with 5,864 images for training and the other 100 species with 5,924 images for testing.
It is generally used to verify the performance of fine-grained retrieval.

Stanford Cars is composed of 16,185 cars images of 196 classes, where the first 98 classes with 8,054 images for training and the other 98 classes with 8,131 images for testing.

The Stanford Online Products dataset has 120,053 images of 22,634 online products (classes) from eBay.com.
We split 11,318 classes with 59,551 images for training and the other 11,316 classes with 60,502 images for testing.
It is generally used to verify the performance of large-scale retrieval.

We also used the noisy label synthesis method in \cite{25}. It can mimic the trait of naturally occurring label noise.
For a more convenient comparison, instead of regenerating a new dataset, we directly use the noise dataset synthesized in \cite{25}.
Among them, the noise is set at 10\%, 20\% and 50\% on both datasets CUB-200-2011 and Cars-196.

\subsection{Ablation Studies}

In this section, we carried out ablation studies on two datasets: CUB-200-2011 and Cars-196, to demonstrate the effectiveness of the Calibrate Proxy structure and Multi-proxy module.

\subsubsection{Validation of the Calibrate Proxy Structure}

In this part, we validate the proposed CP structure by showing the performance of CP-proxy based loss and the role of global center. We also investigate the impact of the settings of global center.

We first use only $S_{ep}$ similarity for loss calculation (this is equivalent to directly using the original Proxy Anchor loss \cite{17_22_26} ($L_{pa}$)).
Then we use both $S_{ep}$ and $S_{em}$ similarity for loss calculation ($L_{c}$).
Finally, in order to optimize the proxy towards the global center, we add constraint loss $L_{mse}$ upon the loss of using both $S_{ep}$ and $S_{em}$ ($L_{c} + L_{mse}$).
The retrieval performance of the two methods is shown in Table.\ref{tab1}.

\begin{table}[h]
  \caption{Validation of the Calibrate Proxy structure}
  \label{tab1}
  \begin{tabular}{llllll}
    \toprule
    Dataset                        & Method            & R@1  & R@2  & R@4  \\
    \midrule
    \multirow{3}{*}{CUB-200-2011 } & $L_{pa}$          & 68.5 & 79.0 & 86.4 \\
                                   & $L_{c}$           & 68.6 & 79.0 & 86.5 \\
                                   & $L_{c} + L_{mse}$ & 69.2 & 79.3 & 86.7 \\
    \midrule
    \multirow{3}{*}{Cars-196 }     & $L_{pa}$          & 86.1 & 91.7 & 95.0 \\
                                   & $L_{c}$           & 86.2 & 91.7 & 95.1 \\
                                   & $L_{c} + L_{mse}$ & 87.0 & 92.3 & 95.2 \\
    \bottomrule
  \end{tabular}
  \centering
\end{table}

From the Table.\ref{tab1}, we found that when $S_{em}$ was added on the basis of $S_{ep}$ ($L_{c}$), its performance was not significantly improved compared to $L_{c}$.
We believe that there is an embedding deviation between the proxy feature and the global center in the proxy-based method, and the two similarities are not in the same spatial distribution.

To validate the effectiveness of Calibrate Proxy structure, we visualize the deviation between proxy and true embedding centers.

\begin{figure}[h]
  \centering
  \includegraphics [width=0.47\textwidth]{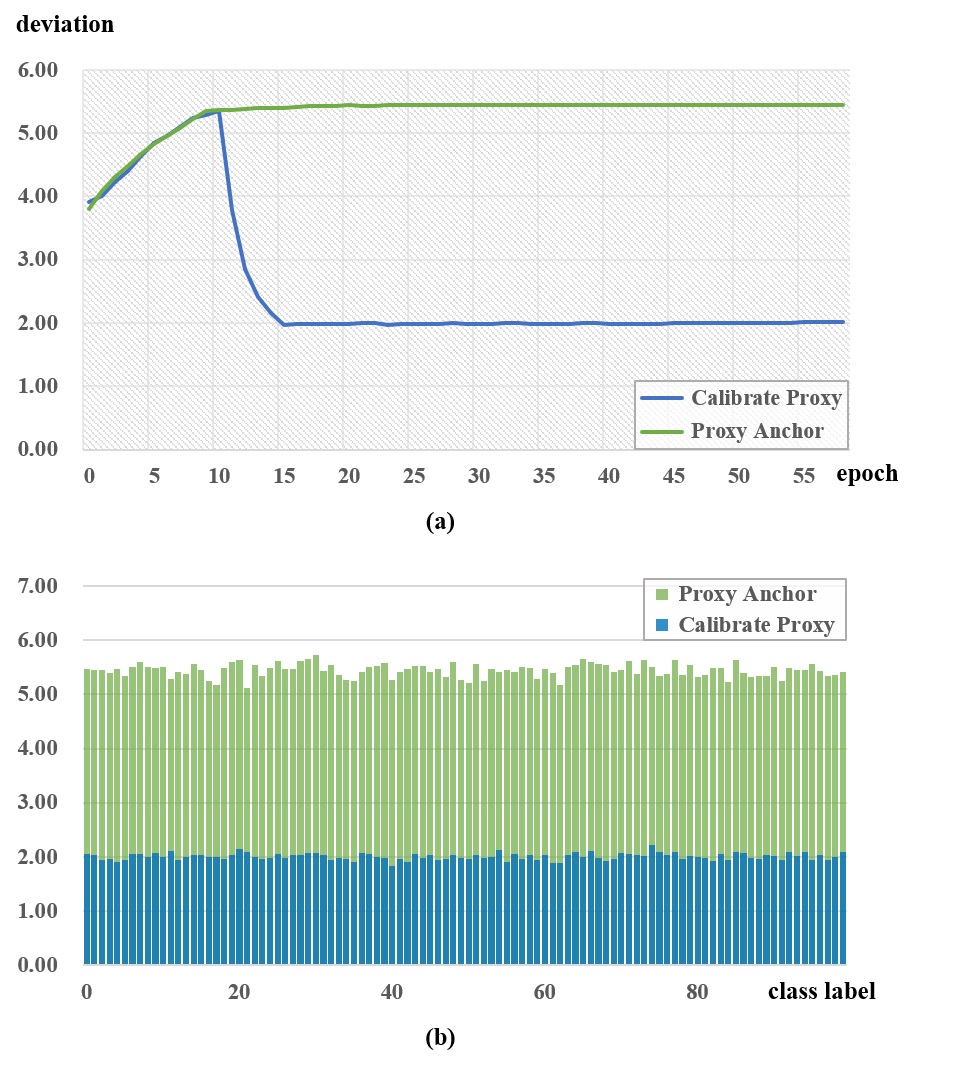}
  \caption{Proxy Anchor loss and Calibrate Proxy Anchor Loss are applied on CUB dataset to show the embedding deviations between the proxy and the real samples. (a) The changes of the embedding deviations with the increase of training epoch. (b) The embedding deviations for each class after network convergence.}
  \label{fig-gap}
\end{figure}

Fig.\ref{fig-gap} (a) shows our CP-proxy anchor loss leads to a quick drop of the embedding deviation at 12-th epoch, where we starts the global center to calculate the proxies.

The deviation between proxy and real sample embedding is calculated as follows,
\begin{equation}
  \label{eq8}
  \begin{aligned}
    d(X_c, P_c) = \| \frac{1}{n_c} \sum_{i=1}^{n_c}{x_i^c} - \frac{1}{n_p} \sum_{j=1}^{n_p}{p_j^c} \Vert
  \end{aligned}
\end{equation}
where $X_c$ denotes all sample embeddings in class $c$, $P_c$ denotes all proxies in class $c$, and $n_c$, $n_p$ denote the number of the sample embeddings and the number of the proxies in class $c$.

We also studied the impact of the global center's capacity and its start epoch on the network performance.

\begin{figure}[h]
  \centering
  \includegraphics [width=0.47\textwidth]{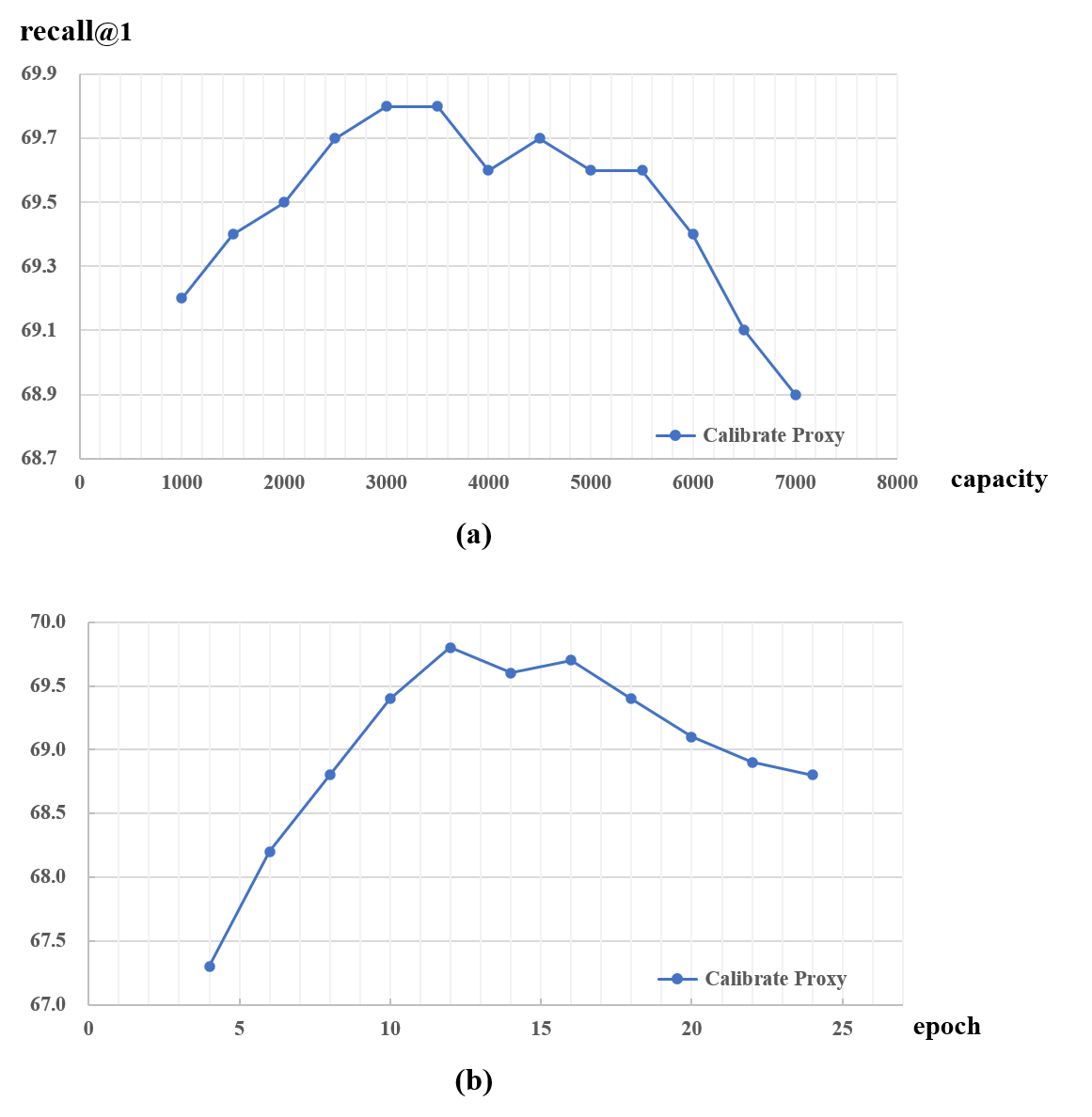}
  \caption{The impact of the global center's capacity and start epoch on performance. (a) Impact of the capacity of the global center. (b) Impact of the start epoch $n_s$ of the global center.}
  \label{fig-global_center}
\end{figure}

As shown in Fig.\ref{fig-global_center} (a) and (b), our proposed method achieves the best peformance at the capacity settings of 3000, 3500 and the start epoch of 12, respectively. Thus, we set the capacity of the global center to 3000 (CUB dataset), storing 30 embeddings per class, i.e., $n_q$=30, and the start epoch $n_s$=12.

To sum up, when we used both $L_{mse}$ and $L_{c}$, the retrieval performance is greatly improved, exceeding the performance of the original Proxy Anchor by 0.9\% on Cars-196.

\subsubsection{Validation of the Multi-proxy Module}

To better express the intra-class feature difference in the embedding space, we appropriately increase the number of proxies in each class to improve the performance.
Therefore, we conduct experiments to analyze the impact of the number of proxies per class.
The results are shown in Table.\ref{tab2}.

\begin{table}[h]
  \caption{Validation of the Multi-proxy module with and without Calibrate Proxy structure}
  \label{tab2}
  \begin{tabular}{lllllllll}
    \toprule
    \multirow{2}{*}{$n_p$} & \multicolumn{3}{c}{CUB-200-2011} & \multicolumn{3}{c}{Cars-196}                             \\
    \cmidrule(r){2-4} \cmidrule(r){5-7}
                           & R@1                              & R@2                          & R@4  & R@1  & R@2  & R@4  \\
    \midrule
    \multicolumn{7}{c}{without Calibrate Proxy structure}                                                                \\
    1                      & 68.4                             & 79.0                         & 86.2 & 86.1 & 91.7 & 95.0 \\
    2                      & 68.7                             & 79.1                         & 86.6 & 86.7 & 92.0 & 95.2 \\
    3                      & 69.0                             & 79.3                         & 86.9 & 87.1 & 92.2 & 95.3 \\
    4                      & 68.9                             & 79.2                         & 86.9 & 87.0 & 92.1 & 95.2 \\
    5                      & 68.9                             & 79.3                         & 87.0 & 87.1 & 92.2 & 95.4 \\
    \midrule
    \multicolumn{7}{c}{with Calibrate Proxy structure}                                                                   \\
    1                      & 69.2                             & 79.3                         & 86.7 & 87.0 & 92.3 & 95.2 \\
    2                      & 69.5                             & 79.6                         & 86.9 & 87.3 & 92.4 & 95.3 \\
    3                      & 69.8                             & 79.7                         & 87.1 & 87.5 & 92.4 & 95.4 \\
    4                      & 69.8                             & 79.7                         & 87.0 & 87.4 & 92.4 & 95.4 \\
    5                      & 69.8                             & 79.6                         & 87.1 & 87.5 & 92.5 & 95.3 \\
    \bottomrule
  \end{tabular}
  \centering
\end{table}

We found that appropriately increasing the number of proxies per class improves performance with or without our proposed Calibrate Proxy structure.
When the number of proxies in each class is 3, the model performance is optimal. Therefore, we set the number of proxies per class to 3.

\subsection{Comparison with the state-of-the-Art}

To verify the performance and noise robustness, in this section we compare our method with the state-of-the-art methods in two sets of experiments.
We first perform experiments on three regular noise-free datasets, and then perform noise-resistance experiments on datasets with different noise ratios.

\subsubsection{Regular Dataset}

We compared our method with the representative methods on the datasets CUB-200-2011 \cite{28}, Cars-196 \cite{30}, Stanford Online Products (SOP) \cite{11}. The comparative methods include Lifted Struct Loss \cite{11}, N-pair Loss \cite{9_20}, ABIER \cite{8395046}, ABE \cite{kim2018attentionbased}, Tuplet Margin Loss \cite{Yu2019}, MS Loss \cite{12}, Circle loss \cite{13}, Proxy-NCA Loss \cite{15}, SoftTriple Loss \cite{16}, Proxy Anchor Loss \cite{17_22_26}, Contrastive Loss \cite{7_23}, CosFace \cite{cosface}, ArcFace \cite{arcface}, Cont.+ XBM \cite{14}, Proxy-NCA++ \cite{Teh2020}, HPL+PA \cite{18}, Mem Vir + PA \cite{Mem}, PS+PA \cite{27}.
We set the $\lambda$ in Eq.(\ref{eq7}) to 1.
The comparison of Recall@k performance is summarized in Table.\ref{tab3}, and the results of MAP@R metrics are summarized in Table.\ref{tab4}. The embedding size of all methods is set to 512.

\begin{table}[h]
  \centering
  \caption{MAP@R performance on CUB-200-2011, Cars-196, and SOP.}
  \label{tab4}
  \begin{tabular}{lccc}
    \toprule
    MAP@R            & CUB-200-2011   & Cars-196       & SOP            \\
    \midrule
    Contrastive      & 26.19          & 25.49          & 44.51          \\
    CosFace          & 26.53          & 26.86          & 46.92          \\
    ArcFace          & 26.45          & 27.22          & 47.41          \\
    MS Loss          & 25.16          & 27.84          & 46.42          \\
    SoftTriple       & 25.64          & 26.06          & 47.35          \\
    Cont. + XBM      & 26.85          & 26.04          & -              \\
    Proxy-NCA++      & 23.53          & 26.02          & 46.56          \\
    Proxy-NCA        & 23.85          & 25.56          & 47.22          \\
    Proxy Anchor     & 26.47          & 27.77          & 47.88          \\
    HPL+PA           & 26.72          & 28.67          & 49.07          \\
    MemVir + PA      & 27.83          & 30.58          & 50.35          \\
    PS + PA          & 28.11          & 29.71          & 47.49          \\
    \textbf{CP + PA} & \textbf{28.34} & \textbf{30.79} & \textbf{50.52} \\
    \bottomrule
  \end{tabular}%
\end{table}%

Specifically, our proposed CP structure with Proxy Anchor loss reaches 69.8\% on the CUB-200-2011, 87.5\% on Cars-196 at Recall@1, and reaches 28.34\% on the CUB-200-2011, 30.79\% on Cars-196 at MAP@R.

\begin{table*}[h]
  \caption{Recall@K(\%) performance on CUB-200-2011, Cars-196 and SOP. Superscript denotes embedding size.}
  \label{tab3}
  \begin{tabular}{lcccccccccccc}
    \toprule
                                            & \multicolumn{4}{c}{CUB-200-2011} & \multicolumn{4}{c}{Cars-196} & \multicolumn{4}{c}{SOP}                                                                                                                                                 \\
    \cmidrule(r){2-5} \cmidrule(r){6-9} \cmidrule(r){10-13}
    \multicolumn{1}{c}{Recall@k}            & 1                                & 2                            & 4                       & 8             & 1             & 2             & 4             & 8             & 1             & 10            & 100           & 1000          \\
    \midrule
    \multicolumn{13}{c}{Pair-based}                                                                                                                                                                                                                                                     \\
    Lifted Struct$^{64}$                    & 43.6                             & 56.6                         & 68.6                    & 79.6          & 53.0          & 65.7          & 76.0          & 84.3          & 62.5          & 80.8          & 91.9          & 97.4          \\
    N-pair$^{64}$                           & 51.0                             & 63.3                         & 74.3                    & 83.2          & 71.1          & 79.7          & 86.5          & 91.6          & 67.7          & 83.8          & 93.0          & 97.8          \\
    ABIER$^{512}$                           & 57.5                             & 68.7                         & 78.3                    & 86.2          & 82.0          & 89.0          & 93.2          & 96.1          & 74.2          & 86.9          & 94.0          & 97.8          \\
    ABE$^{512}$                             & 60.6                             & 71.5                         & 79.8                    & 87.4          & 85.2          & 90.5          & 94.0          & 96.1          & 76.3          & 88.4          & 94.8          & 98.2          \\
    Tuplet Margin$^{512}$                   & 62.5                             & 73.9                         & 83.0                    & 89.4          & 86.3          & 92.3          & \textbf{95.4} & 97.3          & 78.0          & \textbf{91.2} & \textbf{96.7} & \textbf{99.0} \\
    Multi-Similarity$^{512}$                & 65.7                             & 77.0                         & 86.3                    & 91.2          & 84.1          & 90.4          & 94.0          & 96.5          & 78.2          & 90.5          & 96.0          & 98.7          \\
    Circle loss$^{512}$                     & 66.7                             & 77.4                         & 86.2                    & 91.2          & 83.4          & 89.8          & 94.1          & 96.5          & 78.3          & 90.5          & 96.1          & 98.6          \\
    \midrule
    \multicolumn{13}{c}{Proxy-based}                                                                                                                                                                                                                                                    \\
    Proxy-NCA$^{64}$                        & 49.2                             & 61.9                         & 67.9                    & 72.4          & 73.2          & 82.4          & 86.4          & 87.8          & 73.7          & -             & -             & -             \\
    CP + Proxy-NCA$^{64}$                   & 57.1                             & 67.3                         & 72.8                    & 79.9          & 77.8          & 86.1          & 89.7          & 90.8          & 74.7          & 87.3          & 94.1          & 97.4          \\
    SoftTriple$^{512}$                      & 65.4                             & 76.4                         & 84.5                    & 90.4          & 84.5          & 90.7          & 94.5          & 96.9          & 78.3          & 90.3          & 95.9          & -             \\
    CP + SoftTriple$^{512}$                 & 67.6                             & 78.2                         & 85.9                    & 91.6          & 85.3          & 91.3          & 94.9          & 97.1          & 78.8          & 90.7          & 95.9          & 98.5          \\
    Proxy Anchor$^{512}$                    & 68.4                             & 79.2                         & 86.8                    & 91.6          & 86.1          & 91.7          & 95.0          & 97.3          & 79.1          & 90.8          & 96.2          & 98.7          \\
    \textbf{CP + Proxy Anchor Loss}$^{512}$ & \textbf{69.8}                    & \textbf{79.7}                & \textbf{87.1}           & \textbf{92.1} & \textbf{87.5} & \textbf{92.4} & \textbf{95.4} & \textbf{97.6} & \textbf{79.7} & \textbf{91.2} & 96.4          & 98.9          \\
    \bottomrule
  \end{tabular}
  \centering
\end{table*}

\subsubsection{Dataset with Label Noisy}

In this section, we select the state-of-the-art methods in noise-resistance metric learning and classic methods in typical deep metric learning.
The comparison is made on the CUB-200-2011 and Cars-196 with noise ratios of 10\%, 20\% and 50\% respectively.
The comparison results are shown in Table.\ref{tab5}.

\begin{table}[h]
  \caption{Recall@K(\%) performance on CUB-200-2011 and Cars-196 with noisy label. Superscript denotes embedding size, no superscript method set to 512.}
  \label{tab5}
  \begin{tabular}{lccccccccc}
    \toprule
                                   & \multicolumn{3}{c}{CUB-200-2011} & \multicolumn{3}{c}{Cars-196}                                                                 \\
    \cmidrule(r){2-4} \cmidrule(r){5-7}
    \multicolumn{1}{c}{Noisy Rate} & 10\%                             & 20\%                         & 50\%          & 10\%          & 20\%          & 50\%          \\
    \midrule
    \multicolumn{7}{c}{Pair-based}                                                                                                                                   \\
    Circle Loss                    & 47.5                             & 45.3                         & 13.0          & 71.0          & 56.2          & 15.2          \\
    MS Loss                        & 57.4                             & 54.5                         & 40.7          & 66.3          & 67.1          & 38.2          \\
    Contrastive                    & 51.8                             & 51.5                         & 38.6          & 72.3          & 70.9          & 22.9          \\
    MCL                            & 56.7                             & 50.7                         & 31.2          & 74.2          & 69.2          & 46.9          \\
    MCL+PRISM                      & 58.8                             & 58.7                         & 56.0          & 80.1          & 78.0          & 72.9          \\
    \midrule
    \multicolumn{7}{c}{Proxy-based}                                                                                                                                  \\
    Proxy-NCA$^{64}$               & 47.1                             & 46.6                         & 41.6          & 69.8          & 70.3          & 61.8          \\
    CP + NCA$^{64}$                & 48.9                             & 47.2                         & 43.3          & 72.7          & 71.2          & 64.1          \\
    SoftTriple                     & 51.9                             & 49.1                         & 41.5          & 76.2          & 71.8          & 52.5          \\
    CP + SoftTriple                & 54.6                             & 52.8                         & 48.1          & 79.5          & 74.8          & 66.3          \\
    Proxy Anchor                   & 66.1                             & 62.6                         & 57.4          & 83.6          & 79.1          & 71.7          \\
    \textbf{CP + PA}               & \textbf{67.0}                    & \textbf{64.0}                & \textbf{59.3} & \textbf{85.1} & \textbf{81.4} & \textbf{73.6} \\
    \bottomrule
  \end{tabular}
  \centering
\end{table}

We noticed that the current noise-resistance metric learning method needs a specific denoising procedures or parameter when the noise ratio is known to achieve better performance.
In our method, no additional parameters are set for noisy datasets, which is more in line with practical scenarios.

\section{Conclusion}

In this work, we propose a robust Calibrate Proxy (CP) structure, which aims to compensate for the inadequacy of the proxy in expressing class features using real and extensive sample information.
The proposed CP loss requires a Calibrate Proxy structure to be integrated with a CNN network for image retrieval.
This structure increases a global center on the basis of traditional proxy structure, and enables multiple proxy module.
The CP loss combines the traditional sample-proxy loss with a newly-designed class-proxy loss, which calculate the similarity between samples and the global center to bridge the embedding deviation between proxy and data sample.
We further add a constraint to calibrate proxy optimization towards the global center.
In addition, to deal with the variation and regular aggregation of features within the same class, we assign multiple proxies to each class, to optimize the samples toward a closer proxy.
The results show that our method achieves the state-of-the-art performance on both regular and noisy datasets.

% Use \bibliography{yourbibfile} instead or the References section will not appear in your paper
% \nobibliography{aaai23}

\section*{Acknowledgment}

% This work is supported by the National Natural Science Foundation of China \(Grant No.61972240\).
This work is supported by the National Natural Science Foundation of China under Grant 61972240,
and Science and Technology Commission of Shanghai Municipality under Grant 20050501900.

\bibliographystyle{IEEEtran}

\bibliography{refer}

% \newpage

% your own custom macro containing the $\backslash${\tt{includegraphics}} command to make things


%   Use $\backslash${\tt{begin\{IEEEbiography\}}} and then for the 1st argument use $\backslash${\tt{includegraphics}} to declare and link the author photo.
%   Use the author name as the 3rd argument followed by the biography text.
% \end{IEEEbiography}


%   Use $\backslash${\tt{begin\{IEEEbiographynophoto\}}} and the author name as the argument followed by the biography text.

\vfill

\end{document}